\documentclass[aps,prl,showpacs,twocolumn,floats,epsfig,pdflatex]{revtex4}
\usepackage{epsfig}
\usepackage{amsmath}
\usepackage{amsfonts}
\usepackage{graphicx}
\usepackage{amssymb}
\usepackage{amsbsy}

\begin{document}

\title {Non-equilibrium phonon dynamics in trapped ion systems}

\author{T. Dutta $^{(1)}$, M. Mukherjee $^{(1)}$, and K. Sengupta $^{(1,2)}$}
\affiliation{$^{(1)}$Raman Center, Indian Association for the
Cultivation of Science, Jadavpur, Kolkata-700032, India.\\
$^{(2)}$ Theoretical Physics Department, Indian Association for the
Cultivation of Science, Jadavpur, Kolkata-700032, India.}

\date{\today}

\begin{abstract}

We propose a concrete experiment to probe the non-equilibrium local
dynamics of the one-dimensional Bose-Hubbard model using a trapped
ion system consisting of a linear chain of few ${\rm Ba}^+$ ions
prepared in a state of transverse motional mode which corresponds to
a fixed number of phonons per ion. These phonons are well-known to
be described by an effective Bose-Hubbard model. We propose a
protocol which leads to a sudden local sign reversal of the on-site
interaction strength of this Hubbard model at one of the sites and
demonstrate that the subsequent non-equilibrium dynamics of the
model can be experimentally probed by measuring the time-dependent
phonon number in a specific motional state of the Ba$^{+}$ ions. We
back our experimental proposal with exact numerical calculation of
the dynamics of a Bose-Hubbard model subsequent to a local quench.

\end{abstract}


\maketitle

The possibility of the use of systems of ultracold atoms and trapped
ions to emulate theoretical models of strongly correlated condensed
matter systems has been intensely investigated in recent years
\cite{bloch1,zoller1,fermion1,rev1}. One of the first examples in
this regard constituted the realization of the Bose-Hubbard model in
optical lattice systems and characterization of the
superfluid-insulator transition of the model via measurement of
momentum distribution of the bosons \cite{bloch1}. More recently,
there has been several theoretical proposals of emulating various
condensed matter models in trapped ion systems
\cite{duan1,cirac1,cirac2,maciej1}.
The main advantage of the trapped ion emulators over their ultracold
atom counterparts is that they have relatively large separation
between individual ions compared to that between lattice sites of
cold atoms in an optical lattice \cite{rev1}. This allows optical
measurement of the properties of the vibrational states of
individual ions and thus enables one to address the local properties
of the underlying strongly correlated model emulated by these
systems. Also, these systems provide unique opportunity to study the
Bose-Hubbard model in presence of an attractive on-site interaction.

In recent years, it has also been realized that ultracold atomic
systems, which are probably the best examples of experimental
realization of closed quantum systems, provide an unique opportunity
for studying the non-equilibrium dynamics of their constituent atoms
near a quantum critical point \cite{bakr1,simon1}. This has led to a
plethora of theoretical studies of such dynamics \cite{rev2}.
However, in all of these experiments, it has not been possible to
achieve protocols which takes the system out of equilibrium by a
{\it local} change of system parameters. This is mainly due to
several experimental difficulties associated with addressing
individual atoms in an optical lattice setup. Consequently, in most
of the theoretical studies of non-equilibrium dynamics undertaken on
these systems so far \cite{collath1,ehud1,sengupta1}, the effect of
a global time-dependent ramp of a system parameter has been studied
in details; no attempts have been made to study the response of the
system to, for example, a local ramp. Trapped ion emulators provide
us an ideal test bed for studying such dynamics.

In this work, we demonstrate, via proposal of a concrete experiment,
that a system of one-dimensional (1D) trapped ions provide a
requisite experimental setup to study the non-equilibrium dynamics
of a finite-sized 1D  Bose-Hubbard model subsequent to a local
quench. The key differences between our proposal and the
experimental or theoretical studies carried out on cold-atom system
earlier are two fold. First, our proposed setup provides an
implementation of a local quench which changes the on-site
interaction of the emulated model at a given ion site (or
equivalently at a lattice site for phonons). Second, it allows
optical measurement of local physical quantities such as the time
variation of projection of the phonon (boson) number into a motional
state at a given site and hence provide us with direct information
regarding {\it local} phonon (boson) number variation during
non-equilibrium dynamics subsequent to the quench. We provide a
detailed description of the proposed experimental setup and chart
out the parameter space in which such an experiment can be carried
out without violating the phonon number conserving approximation. We
also back our experimental proposal by an exact numerical study of
the non-equilibrium dynamics of the finite-size Bose-Hubbard model
emulated by the trapped ions. We compute the time evolution of the
local boson (phonon) number density $n_i$ subsequent to a sudden
local sign change of the interaction ($U_i \to -U_i$), show that the
variation of $n_i$ can be observed experimentally by measuring the
time dependence of the motional ground state phonon occupation
number at that site ($n_{i0}$), and demonstrate that the amplitude
of the quantum oscillations of $n_{i0}$ is maximal when $J/U_i$ lies
in the critical (crossover for finite-size system) region. We note
that dynamical properties of the Bose-Hubbard model emulated by an
ionic system subsequent to a local quench has not been studied
either experimentally or theoretically so far; our proposed
experiment and the supporting theoretical analysis therefore
constitute a significant extension of our understanding of local
non-equilibrium dynamics of correlated boson systems emulated by
trapped ions.

\begin{figure}
\rotatebox{0}{\includegraphics*[width=\linewidth]{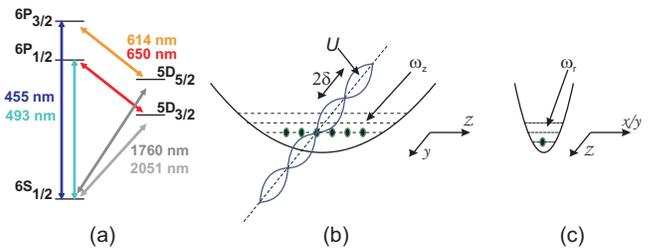}}
\caption{(Color online) (a) Relevant atomic levels of Ba$^+$ ions.
The Zeeman sublevels are not shown for clarity. (b) Schematic
representation of the trapped ion chain. (c) The harmonic oscillator
levels in the radial trap.} \label{Fig1}
\end{figure}

The experimental setup which we propose in the present work
constitutes a linear chain of few Barium ions as shown schematically
in middle and right panels of Fig.\ \ref{Fig1}. The relevant energy
level diagram of a singly charged barium ion is shown in the left panel of Fig.\
\ref{Fig1}. We note here that most of our observations hold
regardless of the nature of the used ions as long as they can
undergo side-band cooling. For preparing the experimental setup, we
trap these ions in a linear trap operated at $15~$MHz
radio-frequency and with the trap stability parameter $q \sim 0.42$
used for the radial confinement. This will generate a
pseudo-potential in the radial plane corresponding to ion
oscillation frequency $\omega_x \simeq 2.25~$MHz. The confinement in
the axial direction is only by DC voltages applied to the endcap
electrodes. This can be made shallow so that the axial frequency is
$\sim 180~$kHz and the mean distance between the ions can be made of
the order to $20~\mu$m~\cite{James}. These parameters will lead to a
tunneling strength of $J \simeq 0.55$~kHz. This in turn leads to
$\beta_{x}=2 J/\omega_x$, the ratio between the coulomb interaction
and the trapping potential, to be $\simeq 5\times10^{-4}$. In a
linear trap like the one presented here, the inter ionic distance in
the Coulomb crystal varies along the chain; however, for the present
purpose this has been neglected as $\beta_{x}\ll1$.

Next, we consider the on-site interaction for the phonons. For this,
we note that for Doppler cooling of the ions, we use a diode laser
at $493$~nm. In addition, the side band cooling can be performed by
driving the red-sideband near the S-D transition frequency at
$2051~$nm. In order to generate the on-site interaction, we use an
off-resonant standing wave laser at $300~$nm which results in a
shift of the radial frequency $\omega_x$. This leads to on site
interaction term $U= 2 (-1)^{\delta} F \eta_{x}^4$, where $F$ is the
strength of the dipole interaction provided by the standing wave,
and $\delta=0(1)$ corresponds to ions being trapped at the
maxima~(minima) of the trapping potential. For $F\sim \omega_{x}$
and $\eta_{x}^{4}=5.26\times10^{-5}$, we find a typical $U= 235$ Hz
which leads to $J/U\sim 2$ . Note that since $J,U \ll \omega_x$, the
number of phonons are conserved. We stress that it is possible to
tune $J/U$ to as low as $0.09$ by suitably choosing $F\sim 25
\omega_{x}$ with $300$~nm standing wave and the radial frequency
shifts to $\omega_{x}\rightarrow 0.52\omega_{x}$ ($1.31\omega_{x}$)
for repulsive (attractive) interactions; the system still remains in
the phonon number conserving regime where
$F\eta_{x}^{2}<<\omega_{x}$. The standing wave pattern will be made
by on-to-one focussing on each ion which eventually allows
individual addressing.


\begin{figure*}
\rotatebox{0}{\includegraphics*[width=\linewidth]{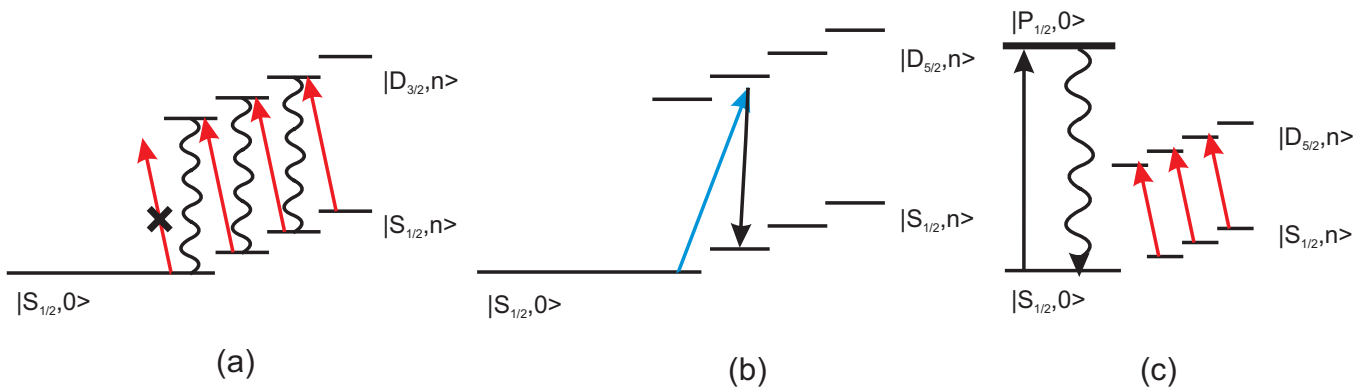}}
\caption{(Color online) A sequence of laser pulses that are needed
to be driven in order to implement site specific quenching. (a)
Sideband cooling laser applied continuously in order to prepare the
1D lattice in the ground state of the system. (b) A Raman PI pulse
is applied in order to prepare the lattice in an initial near
Mott-Insulator state. This pulse is applied to all the sites. (c)
This is the final step after the quench pulse has been applied as
described in the text. In this step shelving of all other states
apart from the motional ground state is performed by a red sideband
(RSB) PI pulse.} \label{Fig2}
\end{figure*}

To implement a local quench on the prepared chain of Doppler cooled
ions, we prepare a series of steps as shown in Fig.\ \ref{Fig2}. The
ion system, after being side band cooled to the motional ground
state, is first prepared for simulation [step(a)]. A Raman type PI
pulse with one blue sideband and one carrier photon applied to all
ions then produces a state with one phonon each at all the sites
[step (b)]. It is well-known, that the low-energy Hamiltonian of
these phonons can be represented by the Bose-Hubbard model given by
\cite{cirac1}
\begin{eqnarray}
H &=& J \sum_{\langle ij\rangle} (b_i^{\dagger} b_j +{\rm h.c.})
+ U \sum_i {\hat n}_i({\hat n}_i -1) + \omega_x \sum_i {\hat
n_i}, \label{ham1} \nonumber\\
\end{eqnarray}
where $b_i$ denotes the annihilation operator of the bosons
(phonons) at site $i$ and ${\hat n}_i= b_i^{\dagger} b_i$ is the
local density operator. Note that Eq.\ \ref{ham1} is identical to
the Bose-Hubbard Hamiltonian in optical lattices with $U/2 \to U$
\cite{cirac1}. To emulate the effect of the quench of $U$ at a given
site, the applied single-site interaction produced by the standing
wave is suddenly quenched by changing the phase of the wave by means
of a peizo actuator with frequency $\omega_0 \sim 50-100$KHz. The
frequency $\omega_0$ is chosen such that it keeps the system in the
phonon number conserving regime ($\omega_0 \ll \omega_x$) and that
it is fast compared to typical energy scales of the underlying
Bose-Hubbard Hamiltonian ($\omega_0 \gg J,U$).  This leads to the
implementation of a local quench protocol. We stress that the
ion-trap system provides a large enough window where quench dynamics
of its underlying Bose-Hubbard Hamiltonian can be studied accurately
within the number conserving approximation. In the rest of this
paper, we restrict ourselves to this regime where $\omega_x \gg
\omega_0 \gg J,U$.


After the implementation of the local-quench protocol, we aim to
measure the subsequent phonon distribution as a function of time. To
this end, we note that the time dependence of the phonon
distribution is reflected in the time-dependent local population of
the phonons $n_{i0}(t)$ at a specified site ($i$) in a specified
motional state, namely the ground state denoted by the index $0$ in
this case. Thus we aim to measure the time-dependent population of
the state $|S_{\frac{1}{2}},0\rangle$ after the quench has been
performed. In order to measure this population, a red-sideband PI pulse
between $S_{\frac{1}{2}}$ and $D_{\frac{5}{2}}$ states are applied
so as to shelve all other motional state population except that of
$|S_{\frac{1}{2}},0\rangle$. Note that the state
$|S_{\frac{1}{2}},0\rangle$ is decoupled from this transition.
Therefore a $S-P$ fluorescence count will give the population in the
motional ground state [step (c)]. The photon count rates can be
measured within a time window of $1$ second which is much larger
than the life time of the excited state $P_{1/2}$ ($7.8$~ns). The
total emitted photon count can be expressed as
\begin{eqnarray}
R_i(t) &=& \langle R\rangle n_{i0}(t) = n_{i0}(t) f \omega\Gamma
Q_{e}Q_{o}/2, \label{pcount}
\end{eqnarray}
where $\langle R \rangle$ is the mean fluorescence photon count
rate; $f=0.73$ is the branching ratio between $P_{1/2}$ and
$S_{1/2}$ state,
$\omega=\Omega/4\pi=1/2[1-\sqrt{1-NA^{2}}]=0.04$~(where $NA=0.4$ is
the numerical aperture of a fluorescence collection Halo lens) is
the fractional solid angle of detection, $Q_{e}$ is the quantum
efficiency of a photomultiplier tube~(PMT) detector which can be
estimated to be about $50$\% at wavelength of $493$~nm and
$Q_{o}=0.1$ is the overall loss factor in collecting the
fluorescence photons in the experimental setup~\cite{Rott}. Such a
fluorescence count measurement as function of time therefore yields
the requisite information on the time-dependent non-equilibrium
population distribution of the phonons subsequent to a local quench.
In what follows, we shall provide a theoretical estimate of this
number for various parameter regime of the Bose-Hubbard model.

\begin{figure}
\rotatebox{0}{\includegraphics*[width=\linewidth]{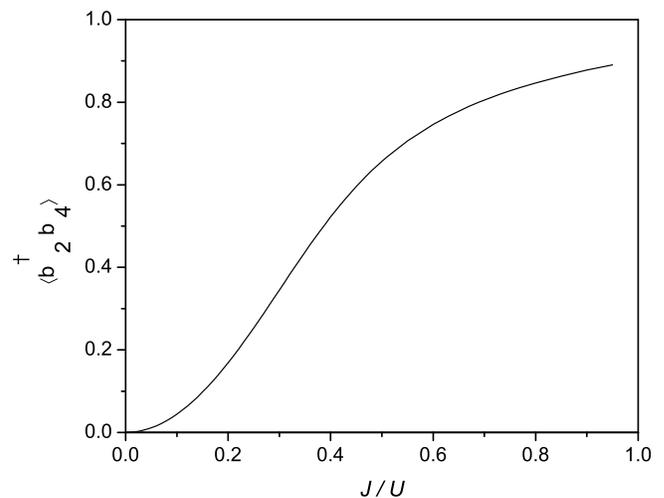}}
\caption{Correlation function of the bosons as a function of $J/U$
indicating a finite-sized crossover from the Mott to the superfluid
phase.} \label{fig3}
\end{figure}
\begin{figure*}
\includegraphics[width=\linewidth]{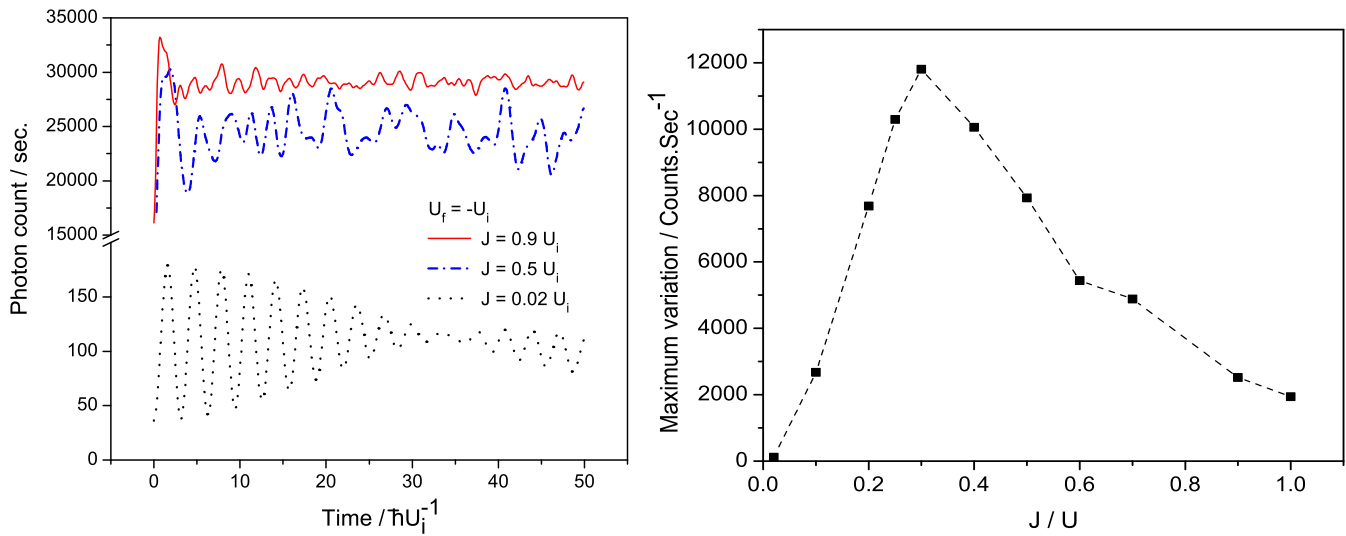}
\caption{(Color online) Left panel: Count rate of
photons detected per second as a function of time (in units of
on-site interaction strength). Right panel: Maximum variation of
count rate of photons detected per sec. at different initial values
of $J/U$ before the applied quench.} \label{fig4}
\end{figure*}

To obtain a theoretical understanding of the dynamics of the
Bose-Hubbard model when the interaction parameter is quenched from
$U_i$ to $U_f$, we use exact diagonalization to obtain the ground
state of the system for $U=U_i$ and all states and energies of the
system at $U=U_f$. Since the total phonon number of the system is
conserved, we restrict ourselves to those states in the Hilbert
space which has a fixed number of total phonons (bosons). In the
rest of this paper, we take this total number of phonons to be equal
to the number of sites. We denote the initial ground state of the
model by $|\psi_G\rangle$ and the energy eigenstates and eigenvalues
at $U=U_f$ after the quench by $|\alpha\rangle$ and $E_{\alpha}$
respectively. The wavefunction of the system at a time $t$ after the
quench at $t=0$ can then be found by solving the Schrodinger
equation $i \hbar
\partial_t |\psi(t)\rangle = H |\psi(t)\rangle$ with the initial
condition $|\psi(t=0)\rangle=|\psi_G\rangle$. Expanding the
wavefunction $|\psi(t)\rangle = \sum_{\alpha} c_{\alpha}(t)
|\alpha\rangle$, one finds
\begin{eqnarray}
|\psi(t)\rangle &=& \sum_\alpha c_{\alpha}^0 e^{-i E_{\alpha}
t/\hbar} |\alpha\rangle, \quad c_{\alpha}^0= \langle
\alpha|\psi_G\rangle. \label{coeff}
\end{eqnarray}
The time evolution of any operator ${\hat O}_i$, where $i$ is the
site index, can be obtained from this wavefunction as $\langle {\hat
O}_i\rangle(t) = \sum_{\alpha,\beta} c_{\alpha}^{0\,\ast}c_{\beta}^0
e^{i(E_{\alpha}-E_{\beta})t/\hbar} \langle \alpha|{\hat
O}_i|\beta\rangle$. In the rest of this work, we shall focus on the
time evolution of the projection of the boson number density on the
motional ground state of the phonons which corresponds to ${\hat O}_i
= {\hat n}_{0i}$ and is given by
\begin{eqnarray}
n_{i0}(t) &=& \sum_{\alpha,\beta,k_0}
c_{\alpha}^{0\,\ast}c_{\beta}^0 \cos[(E_{\alpha}-E_{\beta})t/\hbar]
d^i_{\alpha k_0} d^i_{k_0 \beta},
\end{eqnarray}
where $d^i_{\alpha k_0} = \langle \alpha|\Psi_{k_0}^i\rangle$ is the
projection of the energy eigenstate $|\alpha\rangle$ onto the state
$|\Psi_{k_0}^i\rangle$ which has zero phonons at the i$^{\rm th}$
site, and the sum over $k_0$ indicates a sum over all such states in
the restricted Hilbert space.



The equilibrium phase diagram of the Hubbard model as a function of
$J/U$, as obtained by exact diagonalization is shown in Fig.\
\ref{fig3} where the order parameter correlation $\Delta_{i,j}=
\langle b_i^{\dagger}b_j \rangle$ is plotted as a function of $J/U$.
We find that the plot demonstrates a gradual finite-size crossover
from the Mott phase ($\Delta=0$) at small $J/U$ to a superfluid
phase as $J/U$ is increased. To study the effect of local quench, we
now prepare the system in its ground states with several different
$U$, switch $U_i \to -U_i$ at the i$^{\rm th}$ site with the peizo
actuator, and measure the photon count $R_i(t)$ which reflects
$n_{i0}(t)$ as a function of time (Eq.\ \ref{pcount}). The resulting
photon number fluctuation as a function of time after the applied
quench is shown in the left panel of Fig.\ \ref{fig4}. We find, in
accordance with earlier theoretical predictions in Refs.\
\cite{sengupta2,sengupta3}, that the amplitude of oscillations of
the phonon distribution is maximal when $U_i$ is in the crossover
region as shown in the right panel of Fig.\ \ref{fig4}. Further, we
find that the response of the system to the quench is minimal in the
Mott phase since $n_i$ approximately commutes with $H$ for small
$J/U$ (the commutation is exact for $J=0$) and hence can not change
appreciably as a function of time. From Fig.\ \ref{fig4}, we
estimate that it is feasible to detect a variation of about $30$\%
in the photon count rate (in a count rate of about $24,000$ photons
per sec) for different interaction times in the crossover
regime~\cite{Dubin}. This is well within the present experimental
reach.

In conclusion, we have suggested an experimental proposal of
detecting time evolution of the phonons in a motional ground state
of phonons in an ion trap system subsequent to the {\it local} change
in the effective phonon interaction in one of the ion sites. Our
work shows that such an experiment would shed light on the
properties of the local dynamics of the Bose-Hubbard model obeyed by
these phonons which has not been experimentally studied before. Our
supporting theoretical analysis can be easily generalized to initial
thermal states of the bosons, to quenches with finite ramp rates,
and to other models emulated by these systems.

KS and MM thanks DST-CSIR India for support through grant nos.
SR/S2/CMP-001/2009 and SR/S2/LOP-0024/2007 respectively. TD thanks
CSIR for the financial support.

\end{document}